\documentclass{elsart}

\begin{document}

\begin{frontmatter}

\title{Comment on {\it Conservative force fields in nonextensive kinetic theory }}

\author{F.A. da Costa\corauthref{cor1}} and
\ead{fcosta@dfte.ufrn.br}
\author{M.D.S. de Meneses}
\corauth[cor1]{Corresponding Author}
\ead{mdurval@dfte.ufrn.br}
\address{Departamento de F\'{\i}sica Te\'orica e Experimental \\
Universidade Federal do Rio Grande do Norte \\
Caixa Postal 1641 \\
59072-970, Natal, RN, Brazil}

\begin{abstract}
We discuss an improper application  of the Maxwell-Boltzmann theory to a very unrealistic model. The authors of the original paper claim that a generalized theory would solve a problem that really does not exists. This Comment was submited to Physica A and was not accepted as it is.  
\end{abstract}

\end{frontmatter}


In a recent paper Lima et al. \cite{lima} proposed to generalize the expression
that describes how the density of an ideal gas varies with the height $z$,
measured from the planet surface, in an {\it
isothermal} planetary atmosphere. It is also assumed that the gravitational field
is constant in the $z$-direction, so that the potential energy is simply $U(z) = mgz$.
These simplified hypotheses lead to well-known barometric formula

\begin{equation} \label{boltz}
                               \rho(z) = \rho_0 \exp \left[ -\frac{mgz}{k_B T} \right] ~,
\end{equation}

\noindent
where $\rho_0$ is the gas density at the planet surface level ($z= 0$), $m$ is
the gas mass, $k_B$ represents the Boltzmann constant and $T$ is the absolute temperature.

In the wake of Tsallis \cite{tsallis} non-extensive $q$-Thermostatistics Lima et al. found Eq. (24) presented
in their paper \cite{lima}. These authors claim that for suitable values of $q$ their expression would solve
the problem of having an atmosphere that would extend to infinite. Actually, it is well
known from elementary physics that there is really no such problem with respect to
the Boltzmann distribution which leads to Eq. (\ref{boltz}). Any book of physics for undergraduate
students explains that there are {\it two fundamental hypotheses} behind Eq. (\ref{boltz}), namely,
that {\bf both} $g$ {\bf  and} $T$ are constant as $z$ varies from $0$ to $\infty$. For typical
heights $z \approx 10$ to $100 km$ the error in $g$ is of the order of $1\%$,  so there is not
a serious problem to assume a constant $g$. On the other hand, it is empirically known that the temperature falls
approximately $5K$ as we go $1 km$ higher, if we start from the earth surface. Thus one notes that as $z$ goes from $0$
up to $z=25 km$ the {\it isothermal} condition is {\it \bf violated}, and we could not apply
an equilibrium Boltzmann distribution to deduce Eq. (\ref{boltz}).
On this basis, the Table 1 presented in \cite{lima} is completely void of significance (we think there is problably a
misprinting for the $z_{max}$ of Oxygen when $q=0.8$). Also, that Table gives figures for the Hydrogen and
Oxygen gases. If the authors had considered the most abundant gases in our atmosphere, they would have
found a somewhat better results {\it considering} $q=0.7$, namely, $z_{max} = 18 km$, and $29 km$ for
$CO_2$ and $N_2$, respectively. However, we were not able to find any criterion in their paper about the 
right choice for the value of the parameter $q$. 

For pedagogical reasons we thing that any discussion relative to distances in this sort of problem should begin 
by considering the natural length scale for this problem, namely,

$$ \xi  = \frac{k_B T}{mg}  . $$

Taking, for instance, $T = 300 K$ we can easily obtain $\xi_{H_2} \approx 100 km$, whereas $\xi \approx 10 km$ for
$O_2$, $N_2$ and $CO_2$.

In summary, contrary to the authors claim, there is really no problem with the Maxwell-Boltzmann distribution
in the present context. The problem we are faced with in this case lies in the model which is {\it completely
inappropriated} to discuss the question raised Lima et el. \cite{lima}.

\end{document}